\documentclass[pra,showpacs,showkeys,preprint]{revtex4}
\usepackage[utf8]{inputenc}
\usepackage{graphicx}
\usepackage{amssymb}
\pdfoutput=1
\begin{document}
\title{Mapping the charge-dyon system into the position-dependent effective mass background via Pauli equation}

\author{Anderson L. de Jesus\footnote{Corresponding author} and Alexandre G. M. Schmidt}
\email{agmschmidt@id.uff.br, Telephone: (+55)
24 3076 8935, FAX: (+55) 24 3076 8870} \affiliation{Departamento de F\'isica de Volta Redonda,\\ Instituto de Ci\^encias Exatas --- Universidade Federal Fluminense, \\ R.  Des. Ellis Hermydio Figueira, 783, Bloco C, Volta Redonda RJ, CEP 27213-145, Brazil}

\date{\today}

\begin{abstract}
This work aims to reproduce a quantum system composed of a charged spin - $1/2$ fermion interacting with a dyon with an opposite electrical charge (charge-dyon system), utilizing a position-dependent effective mass (PDM) background in the non-relativistic regime via the PDM free Pauli equation. To investigate whether there is a PDM quantum system with the same physics (analogous model) that a charge-dyon system (target system), we resort to the PDM free Pauli equation itself. We proceed with replacing the exact bi-spinor of the target system into this equation, obtaining an uncoupled system of non-linear partial differential equations for the mass distribution $M$. We were able to solve them numerically for $M$ considering a radial dependence only, i.e., $M=M(r)$, fixing $\theta_0$, and considering specific values of $\mu$ and $m$ satisfying a certain condition. We present the solutions graphically, and from them, we determine the respective effective potentials, which actually represent our analogous models. We study the mapping for eigenvalues starting from the minimal value $j = \mu - 1/2$.
\end{abstract}

%\pacs{xyz}
\keywords{Position-dependent mass systems; Pauli equation; Dyon.}

\maketitle

\section{Introduction}

The study of magnetic monopoles is an interesting and active area of physics. Despite its non-experimental verification, the magnetic monopole is still a subject of investigation by several authors; Dirac in the 1930s explained electric charge quantization due to the presence of magnetic monopoles and his paper \cite{dirac} was probably one of the works that motivated several researchers for the investigation of the theoretical aspects of the physics of the magnetic monopoles. As well as the magnetic monopole --- used here as a synonym of a particle having a magnetic charge ---,  dyon is also a hypothetical particle having electric and magnetic charges simultaneously. Such particle was proposed by  Schwinger in 1969 as a phenomenological alternative to quark model \cite{dyons}. In his article, Schwinger speculates that hadrons can be composed of these dual particles, having fractional electric and magnetic charges and considering the idea of electroweak interaction with its vectorial bosons exchanging electric charge, he postulated the existence of a new vectorial boson $S$ (strong) of unit magnetic charge intermediating charge-exchange process for the dyon. The study of dyons and isolated magnetic monopoles has received some attention over the latest years, and several works on this topic are available in the literature \cite{shnir,moriyasu,rossi,exposed,diatoms, volcanic}. Recently, A. Eriksson and E. Sj\"oqvist carried out a research about monopole field textures in interacting spin systems \cite{monopole textures}. 

On the other hand, quantum systems with position-dependent effective mass (PDM quantum systems) have been studied and used in a great number of works and a considerable interest in this subject has also grown up over the latest years. PDM quantum systems arose initially in the study of transport phenomena in semiconductors of variable, position-dependent chemical composition. The Hamiltonian for PDM quantum systems in its more general form and the ordering problem that it carries were studied by von Roos in 1980s \cite{roos}. Zhu and Kroemer studied in 1983 an abrupt heterojunction between two different semiconductors, proposing in this work a given ordering \cite{Zk}. Mustafa and Mazharimousavi carried out in 2007 an important work about ordering ambiguity via PDM pseudo-momentum operators, proposing a different ordering \cite{mustafa}. Both Mustafa-Mazharimousavi and Zhu-Kroemer ordering are physically allowed by the Dutra and Almeida Test \cite{dutra}. PDM quantum systems have been investigated por Mustafa and Algadhi, resulting in recent works, such as: Position-dependent mass charged particles in magnetic and Aharonov–Bohm flux fields: separability, exact and conditionally exact solvability \cite{mustafa1}; Landau quantization for an electric quadrupole moment of position-dependent mass quantum particles interacting with electromagnetic fields \cite{mustafa2}; Position-dependent mass momentum operator and minimal coupling: point canonical transformation and isospectrality \cite{mustafa3}.

PDM quantum systems were also investigated by Yu, Dong and co-authors, resulting in a series of works, such as: solutions of the PDM Schr\"odinger equation for the Morse potential \cite{morse potential}; exactly solvable potentials for the Schr\"odinger equation with spatially dependent mass \cite{solvable}; exact solutions of the PDM Schr\"odinger equation for a hard-core potential \cite{hardcore}; algebraic approach to the position-dependent mass Schr\"odinger equation for a singular oscillator \cite{oscillator}; solution of the Dirac equation with position-dependent mass in a Coulomb and scalar fields in a conical space-time \cite{conical spacetime}. Such systems can also be used to model scattering in abrupt heterostructures \cite{heterostructure}, quantum dots \cite{dots} and in mapping conical spaces \cite{conico pdm}.

\subsubsection{Analogous models}

Analogous models for systems including magnetic monopoles have already been investigated over the last years by several other authors and some of these models were even obtained experimentally. Thus, in the theoretical and experimental aspect there is also plenty of research and the physics of magnetic monopoles was reproduced in spin ice system \cite{spin ice, spin ice 2, spin ice 3}, as well as, more recently, in quantum fields \cite{qf}; synthetic magnetic fields \cite{synth}; creation of a Dirac monopole-antimonopole pair in a spin-1 Bose-Einstein condensate \cite{finlandes} and by the use of metamaterials \cite{metamaterials}. We will list some of these recent works in the table I.
\vspace{1cm}
\begin{table}
\begin{center}
{\footnotesize 
\begin{tabular}{|c|c|c|}
\hline Target systems $\!\!\!$ & Analogous models $\!\!\!$ & Authors / year  $\!\!\!$ \\ 
\hline Dirac monopole-antimonopole $\!\!\!$ & Spin-1 Bose–Einstein condensate $\!\!\!$ & Tiurev \textit{et al.}/2019  \cite{finlandes} $\!\!\!$\\ 
\hline Dirac monopole-antimonopole $\!\!\!$ &  Spin ice $\!\!\!$ & Castelnovo \textit{et al.}/2018 \cite{spin ice} $\!\!\!$\\ 
\hline Dirac monopole $\!\!\!$ &  Spin ice $\!\!\!$ & Revell \textit{et al.}/2013 \cite{spin ice 3} $\!\!\!$\\
\hline Isolated monopole $\!\!\!$ & Metamaterials $\!\!\!$  & Wang \textit{et al.}/2014 \cite{metamaterials} $\!\!\!$\\ 
\hline Isolated monopole $\!\!\!$ & Spin-1 Bose–Einstein condensate $\!\!\!$ & Ray \textit{et al.}/2014 \cite{qf} $\!\!\!$ \\ 
\hline Isolated Dirac monopole $\!\!\!$ & Nanoscopic magnetized needle $\!\!\!$  & Béché \textit{et al.}/ 2013 \cite{exposed} $\!\!\!$ \\ 
\hline Monopole and Dirac strings $\!\!\!$ & Spin ice $\!\!\!$ & Morris \textit{et al.}/ 2009 \cite{spin ice 2}$\!\!\!$\\ 
\hline 
\end{tabular} }
%\vspace{0.1cm}
\caption{Selection of some recent works on analogous models, which reproduce the physics of the magnetic monopole (antimonopole). Their respective authors and year of publication are also presented.}
\end{center}
\end{table}

Considering magnetic monopoles and dyons, the investigation about analogous models for charge-monopole and charge-dyon systems using PDM quantum systems (PDM Schr\"odinger equation) was initially carried out by the authors in \cite{monopole pdm, dyon pdm} where an exact mapping was constructed for both cases. Later the authors carried out an investigation about the mapping between a charged spin- $1/2$ fermion interacting with a monopole and PDM quantum systems via PDM free Pauli equation \cite{monopole pdm pauli}. We summarize in Table II the analogous models already built and investigated by the authors using PDM quantum systems.
\vspace{1cm}
\begin{table}
\begin{center}
{\footnotesize 
\begin{tabular}{|c|c|c|}
\hline Target systems $\!\!\!$ & Analogous model $\!\!\!$ & Authors / year  $\!\!\!$ \\ 
\hline Charge-monopole system $\!\!\!$ & PDM quantum systems$\!\!\!$ & A. Schmidt, A. de Jesus / 2018  \cite{monopole pdm}$\!\!\!$\\ 
\hline Charge-dyon system $\!\!\!$ & PDM quantum systems $\!\!\!$ &  A. de Jesus, A. Schmidt / 2019 \cite{dyon pdm} $\!\!\!$\\ 
\hline Relativistic charge-dyon system $\!\!\!$ & PDM quantum systems $\!\!\!$  &  A. de Jesus, A. Schmidt / 2019 \cite{dyon pdm} $\!\!\!$\\ 
\hline Spin - 1/2 charge-monopole system $\!\!\!$ & PDM quantum systems $\!\!\!$ &  A. de Jesus, A. Schmidt / 2019 \cite{monopole pdm pauli} $\!\!\!$ \\ 
\hline Conical space $\!\!\!$ & PDM quantum systems $\!\!\!$ & A. de Jesus, A. Schmidt / 2019 \cite{conico pdm} $\!\!\!$\\ 
\hline 
\end{tabular} }
%\vspace{0.1cm}
\caption{Summary of the target systems and the analogous model using PDM quantum systems, which were already built and investigated by the authors. The year of publication is also presented.}
\end{center}
\end{table}

It is useful to comment that PDM quantum systems using the Pauli equation are still little explored in the literature. However, a few years ago, the authors investigated quantum systems with spin-orbit interaction via Rashba and Dresselhaus terms utilizing the usual and PDM Pauli equation \cite{rashba pdm}.

In the present work our goal is to make another contribution to this subject, setting up a map between the charge-dyon and a PDM quantum system via PDM free Pauli equation, i.e., we intend to reproduce, utilizing a PDM background, the physics of a constant mass, charged spin-$1/2$ fermion interacting with a dyon, with an opposite electrical charge, at origin (the charge-dyon system is a bound system which was not observed in nature yet). It is important to comment that there are some similarities between the mappings utilizing charge - monopole and charge - dyon  systems, the difference is that the latter involves confluent hypergeometric functions $_1F_1$, as well as the product $eQ$, where $e$ is the electric charge of the spin-$1/2$ fermion and $Q$ is the electrical charge of the dyon (with $eQ < 0$). Both the models will match considering $Q = 0$. In this sense, the present work is a continuation of the previous work developed in reference \cite{monopole pdm pauli}

The outline for our paper is the following: in section II, we present the PDM free Pauli equation and derive the effective potential in the Zhu-Kroemer parametrization. In section III we build the uncoupled system of mapping equations for charge-dyon system, satisfied by a mass distribution $M(r)$ in the non-relativistic case, and solve them numerically for the case where the eigenvalues start from the minimal value $j = \mu - 1/2$. The solutions are presented graphically. Finally, in section IV, we conclude the work.

\section{Pauli free equation in position-dependent effective mass background}

In this section we present the basic structure of PDM quantum systems, starting from the Hamiltonian for these systems, known as von Roos Hamiltonian $H_{roos}$. We can build the PDM Pauli equation for the free case, i.e., in the absence of electromagnetic fields, as we will see. In this work we will use natural units where $\hbar = m_e = c = k_0 = 1$, $e=\sqrt{\alpha}$ (here $\alpha$ is the so-called fine-structure constant, namely $\alpha \approx 1/137$) and we will choose $M(x_i)=2m(x_i)$, where $x_i$ is a spatial coordinate. We sometimes will omit the function arguments for simplicity. The von Roos Hamiltonian $H_{roos}$ for position-dependent effective mass is given by the following expression,
\begin{equation}\label{H0}
H_{roos}=-\frac{1}{2}\left[M^{\alpha}\vec{\nabla}M^{\beta}\vec{\nabla}M^{\gamma}+M^{\gamma}\vec{\nabla}M^{\beta}\vec{\nabla}M^{\alpha}\right] + V_{ext},
\end{equation}
where $\alpha$, $\beta$ and $\gamma$ are the von Roos ambiguity parameters and they obey the constraint $\alpha+\beta+\gamma=-1$. Let us consider here the external potential $V_{ext} = 0$, it makes (\ref{H0}) be a purely kinetic Hamiltonian $H_0$. This Hamiltonian may be written in a form where the ambiguity parameters lie inside an effective potential $U_{eff}$ \cite{quesne}. After some algebra, we can write (\ref{H0}) as,
\begin{equation}\label{H02}
H_0 =-\frac{1}{M}\nabla^2+\frac{1}{M^2}\vec{\nabla}M\cdot\vec{\nabla}+U_{eff},
\end{equation}
this expression will be utilized throughout our work. After eliminating $\gamma$, $U_{eff}$ has the following form in terms of $\alpha$ and $\beta$ parameters,
\begin{equation}\label{pot eff}
U_{eff}=\frac{(\beta+1)}{2}\frac{\nabla^2 M}{M^2}-[\alpha(\alpha+\beta+1)+\beta+1]\frac{(\vec{\nabla}M)^2}{M^3}.
\end{equation}
The free Pauli equation can be obtained from the usual Pauli equation \cite{Cohen}, that is,
\begin{equation}\label{Pauli}
\frac{1}{2m}(\vec{\sigma}\cdot[\vec{P}-e\vec{A}])^2\psi+eV\psi=E\psi,
\end{equation}
where $\vec{P}=-i\hbar\vec{\nabla}$ is the momentum operator, $\vec{A}$ is the vector potential, $V$ is the scalar electric potential and $\vec{\sigma}=\sigma_{x}\hat i+ \sigma_{y}\hat j +\sigma_{z} \hat k$ is the Pauli vector ($\sigma_{x}, \sigma_{y}, \sigma_{z}$ are the Pauli matrices). In the case where $\vec{A}=V= 0$, the equation (\ref{Pauli}) becomes,
\begin{equation}\label{P}
-\frac{1}{2m}(\vec{\sigma}\cdot\vec{\nabla})^2\psi=E\psi
\end{equation}
An important relation involving the Pauli vector, dot and cross product is \cite{Sakurai, Cohen1, Cohen},
\begin{equation}
(\vec{\sigma}\cdot\vec{a})(\vec{\sigma}\cdot\vec{b})=I_{2}(\vec{a}\cdot\vec{b}) + i(\vec{a}\times\vec{b})\cdot\vec{\sigma},
\end{equation}
where $I_{2}$ is the $2\times 2$ identity matrix. Replacing $\vec{a}=\vec{b}=\vec{\nabla}$ in the previous relation, we have $(\vec{\sigma\cdot\vec{\nabla}})^2\psi=I_{2}(\vec{\nabla}\cdot\vec{\nabla}\psi)+i(\vec{\nabla}\times\vec{\nabla}\psi)\cdot\vec{\sigma}=I_{2}\nabla^2\psi$ and the equation (\ref{P}) becomes,
\begin{equation}\label{free}
I_{2}(-\frac{1}{2m}\nabla^2)\psi=I_{2}H_{0}\psi=E\psi
\end{equation}
The equation (\ref{free}) is the so-called free Pauli equation (note that the Pauli matrices $\sigma_i$ drop out in free case). Taking $\psi$ as a bi-spinor (Pauli bi-spinor), the free Pauli equation in matrix form simply reads,
\begin{equation}\label{Pauli PDM}
\left ( \begin{array}{cc}
H_0 & 0 \\
0 & H_0 \\
\end{array} \right )\left ( \begin{array}{c}
\psi_+\\ \psi_-
\end{array} \right)=E \left ( \begin{array}{c}
\psi_+\\ \psi_-
\end{array} \right).
\end{equation}
The component $\psi_+$ is related to spin up and $\psi_-$, to spin down. The equation (\ref{free}) becomes the PDM free Pauli equation if we replace the kinetic operator $H_{0}$ by the von Roos Hamiltonian (\ref{H02}). 
 
To determine the effective potential $U_{eff}$ for a particular system, we need to know the mass distribution $M$ and also to select a given ordering that must obey some physical considerations. In this work we will use the Zhu-Kroemer ordering \cite{Zk} where,
\begin{equation}
\alpha=\gamma = -1/2,\quad\quad\beta=0,
\end{equation}
that is, a parametrization which fulfills the Dutra and Almeida test \cite{dutra}. This choice implies an effective potential $U_{eff}$, given by the following expression,
\begin{equation}\label{effective potential ZK}
U_{eff}=\frac{1}{2}\frac{\nabla^2 M}{M^2}-\frac{3}{4}\frac{(\vec{\nabla}M)^2}{M^3}.
\end{equation}
In the next section we will utilize this effective potential in the construction of our analogous model.

\section{Mapping a non-relativistic charge-dyon system into position dependent effective mass background by the PDM free Pauli equation}

In the present section we will build the exact mapping between charge-dyon and PDM quantum systems via PDM free Pauli equation, that is, we will perform a map between a system composed of a charged fermion of constant mass $m_0=1$ and spin - $1/2$ ---  up or down --- interacting with a dyon located at the origin into a charged spin - $1/2$ fermion interacting with an effective potential $U_{eff}$. In order to obtain such a mapping we replace the exact wavefunction (Pauli bi-spinor) of the charge-dyon system into the PDM free Pauli equation (\ref{Pauli PDM}) and solve it for the mass $M$. The analogous model --- a charged spin - $1/2$ fermion with position-dependent effective mass, which is equivalent to a charged spin - $1/2$ fermion with constant mass interacting with the effective potential (\ref{effective potential ZK}) --- reproduces exactly the behavior of the original charge-dyon system (target system). From now on in our calculations we will utilize spherical coordinates $(r,\theta,\phi)$. Using the Laplacian in spherical coordinates and the effective potential (\ref{effective potential ZK}), the equation (\ref{free}) yields,
\begin{eqnarray}\label{Model equation dyon}
&&-\frac{I_{2}}{M}\left\{ \frac{1}{r^2}\frac{\partial }{\partial r}\left(r^2\frac{\partial \psi}{\partial r}\right) +\frac{1}{r^2\sin\theta} \frac{\partial }{\partial \theta} \left(\sin\theta \frac{\partial \psi}{\partial \theta}\right) + \frac{1}{r^2\sin^2\theta} \frac{\partial^2 \psi} {\partial \phi^2}\right\} + \frac{I_{2}}{M^2} \vec{\nabla}M\cdot \vec{\nabla} \psi \nonumber\\
&& + I_{2}\left( \frac{1}{2M^2}\nabla^2M - \frac{3}{4M^3} \vec{\nabla}M\cdot\vec{\nabla}M\right) \psi = E_N\psi.
\end{eqnarray}
In the charge-dyon system, because of the electric charge $Q$ of the dyon, the energy spectrum is discrete, labelled by a quantum number $N = 0, 1, 2, 3 ...$. The operator $K = \vec{\sigma}\cdot(\vec{L}- \mu \hat{r})$ \cite{shnir} will be very important to deal with the angular part. The operator $L^2$ reads,
\begin{equation}\label{k2 dyon operator}
L^2=-\frac{1}{\sin\theta}\frac{\partial}{\partial\theta}\left(\sin\theta\frac{\partial}{\partial\theta}\right) - \frac{1}{\sin^2\theta}\frac{\partial^2}{\partial\phi^2}+\frac{2i\mu(1-\cos\theta)}{\sin^2\theta}\frac{\partial}{\partial\phi}+\frac{\mu^2(1-\cos\theta)^2}{\sin^2\theta}+\mu^2,
\end{equation}
where $\mu = eg$, being $e$ the electric charge and $g$ the magnetic charge. The Dirac quantization condition implies that $ \mu=eg= n/2$, with $n$ being an integer. An important relation involving $K$, $K^2$ and $L^2$ \cite{shnir, rossi} is,
\begin{equation}\label{K2 operator}
\mu^2 - L^2 = \mu(\vec{\sigma}\cdot\hat{r})-K^2-K,
\end{equation}
this relation will be useful to lead with the angular part of our mapping equations. Let us point out an important class of functions occurring frequently in magnetic monopole quantum theory: such class of functions is composed by the eigenfunctions of the operator $L^2$ and they appear in the angular part of the exact wavefunctions of charged particles interacting with monopoles. These functions are called generalized spherical harmonics or monopole harmonics, \cite{shnir}, 
\begin{equation}\label{harmonico}
Y_{\mu lm}(\theta,\phi)  = 2^m\sqrt{\frac{(2l+1)(l-m)!(l+m)!}{4\pi (l-\mu)!(l+\mu)!}} \frac{P_{l+m}^{(-\mu-m,-\mu+m)}(\cos\theta)\exp[{i(m+\mu)\phi}]}{(1-\cos\theta)^{(\mu+m)/2} (1+\cos\theta)^{(\mu-m)/2}},
\end{equation}
where $P_n^{(a,b)}(\cos\theta)$ are the so-called Jacobi polynomials \cite{rainville, SF}. In the particular case where $\mu = 0$, these functions reduce to standard spherical harmonics \cite{Cohen1}. 

The dyon charge $Q$ will be a multiple of the elementary electric charge $e = \sqrt{\alpha}$, that is, $Q= ne=n\sqrt{\alpha}$ with $n \in \mathbb Z$, so we can write $eQ=n\alpha$, however we will consider only non-positive values for $n$ (bound state). The operator of generalized angular momentum $\vec{J}= \vec{L}+\frac{1}{2}\vec{\sigma}$ has eigenvalues $j$. The spectrum of eigenvalues of the operator of angular momentum $\vec{L}$ of a spinless charge-monopole system starts from the minimal value $l=\mu$, being $l=\mu, \mu + 1, \mu + 2 ...$ and $-l \leq m \leq l$. Introducing the angular momentum according to the standard rule of angular momenta means that the total angular momentum $\vec{J}$ has eigenvalues $j = l \pm 1/2$. Thus, the spectrum of eigenvalues $j$ can start either from the minimal value $j=\mu - 1/2$ or from the minimal value $j=\mu + 1/2$, being $-j \leq m \leq j$. These situations have to be considered separately. However in this paper we will only build the mapping for the case $j = \mu - 1/2$, named type 3 \cite{shnir}.

\subsection{Exact mapping for $j = \mu - 1/2$}

The wavefunctions for the charge-dyon system, which correspond to the eigenvalues of operator $J$ starting from $j = \mu - 1/2$ are,
\begin{equation}\label{spinor dyon 3}
\psi^{(3)}(r, \theta,\phi)=F_{k\ell}^{(3)}(r)\Omega_{\mu,j,m}^{(3)}(\theta,\phi), 
\end{equation}
as mentioned before, sometimes we will omit the function arguments for simplicity. The index (3) indicates "type 3". The radial part is given by the following expression \cite{shnir},
\begin{equation}
F_{k\ell}^{(3)}(r)=2\left(\frac{n\alpha}{N+1}\right)^{3/2}e^{-kr} \;_1F_1(-N;2;2kr),
\end{equation}
where $k= \sqrt{- 2m_0E_N^{(3)}} > 0$, being $m_0$ the constant mass of the charged fermion to be mapped (we will choose $m_0 = m_e = 1$). The function $_1F_1$ is the so-called confluent hypergeometric function and is defined as solution of confluent hypergeometric equation, which can be obtained by singularities from a Funchsian equation with three singular points (for more details, see reference  \cite{rainville}, chapter 7), $N$ is a radial quantum number $N=0,1,2,...$ and $\ell$ is related to the angular quantum number $j$ via,
\begin{equation}
\ell=\sqrt{\left(j+\frac{1}{2}\right)^2-\mu^2},
\end{equation}
the radial solution satisfies the following radial equation \cite{shnir},
\begin{equation}\label{radial dyon 3 type}
\left[\frac{d^2}{dr^2}+\frac{2}{r}\frac{d}{dr}+\frac{2m_0 eQ}{r}\right]F_{k\ell}^{(3)}(r)= - 2m_0E_N^{(3)}F_{k\ell}^{(3)}(r),
\end{equation}
The angular part is given by the following bi-spinor,
\begin{equation}\label{spinor}
\Omega_{\mu,j,m}^{(3)}(\theta,\phi) = \left ( \begin{array}{c}
 -\sqrt{\frac{\mu-m+1/2}{2\mu+1}} Y_{\mu\mu m-1/2}(\theta,\phi)\\ \sqrt{\frac{\mu+m+1/2}{2\mu+1}} Y_{\mu\mu m+1/2}(\theta,\phi)
\end{array} \right),
\end{equation}
which is an eigenspinor of the operators $K$ and $(\vec{\sigma}\cdot\hat{r})$ \cite{shnir}, that is,
\begin{equation}\label{angular dyon 3}
K \Omega_{\mu,j,m}^{(3)}(\theta,\phi)=-\Omega_{\mu,j,m}^{(3)}(\theta,\phi),\quad\quad(\vec{\sigma}\cdot\hat{r}) \Omega_{\mu,j,m}^{(3)}(\theta,\phi)=\Omega_{\mu,j,m}^{(3)}(\theta,\phi),
\end{equation}
We will rewrite, for sake of simplicity, the radial function $F_{k\ell}^{(3)}(r)$ as $F^{(3)}$, the spinor $\Omega_{\mu,j,m}^{(3)}(\theta,\phi)$ as $\Omega^{(3)}$ and its up and down components as $\Omega_{+}^{(3)}$ and $\Omega_{-}^{(3)}$ respectively.

Let us begin by isolating in operator (\ref{k2 dyon operator}) the angular terms, which appear in the Laplacian of equation (\ref{Model equation dyon}). Thus, isolating these two terms and utilizing the relation (\ref{K2 operator}) to eliminate $\mu^2-L^2$, we have,
\begin{equation}\label{L}
\frac{1}{\sin\theta}\frac{\partial}{\partial\theta}\left(\sin\theta\frac{\partial}{\partial\theta}\right) +\frac{1}{\sin^2\theta}\frac{\partial^2}{\partial\phi^2} = \frac{2i\mu(1-\cos\theta)}{\sin^2\theta}\frac{\partial}{\partial\phi}+\frac{\mu^2(1-\cos\theta)^2}{\sin^2\theta}+ \mu(\vec \sigma \cdot\hat r) - K^2 - K,
\end{equation}
using the equations (\ref{angular dyon 3}), we can check the action of the operator (\ref{L}) on the spinor $\Omega^{(3)}$. The result of this action is,  
\begin{equation}\label{L2}
\frac{1}{\sin\theta}\frac{\partial}{\partial\theta}\left(\sin\theta\frac{\partial \Omega^{(3)}}{\partial\theta}\right) +\frac{1}{\sin^2\theta}\frac{\partial^2\Omega^{(3)}}{\partial\phi^2} = \frac{2i\mu(1-\cos\theta)}{\sin^2\theta}\frac{\partial\Omega^{(3)}}{\partial\phi}+\frac{\mu^2(1-\cos\theta)^2}{\sin^2\theta}\Omega^{(3)} + \mu \Omega^{(3)},
\end{equation}
Replacing (\ref{L2}) in equation (\ref{Model equation dyon}) and utilizing the matrix form, we obtain a system of uncoupled equations for each spinor component $\psi_{\pm}$. It reads,
\begin{eqnarray}\label{L3}
&&-\frac{\Omega^{(3)}_{\pm}}{M r^2}\frac{\partial }{\partial r}\left(r^2\frac{\partial F^{(3)}}{\partial r}\right) +\frac{F^{(3)}}{M r^2} \left(-\mu - \frac{2i\mu(1-\cos\theta)}{\sin^2\theta}\frac{\partial}{\partial\phi}-\frac{\mu^2(1-\cos\theta)^2}{\sin^2\theta}\right)\Omega^{(3)}_{\pm} \nonumber\\
&&+ \frac{1}{M^2}\vec{\nabla}M \cdot \vec{\nabla}\psi^{(3)}_{\pm}+  \left(\frac{1}{2M^2}\nabla^2M - \frac{3}{4M^3} \vec{\nabla}M\cdot\vec{\nabla}M\right) F^{(3)} \Omega^{(3)}_{\pm} = E_{N} F^{(3)} \Omega^{(3)}_{\pm}, 
\end{eqnarray}
We can eliminate the radial part in (\ref{L3}) isolating the radial derivatives in equation (\ref{radial dyon 3 type}) and exchanging for this radial part. This operation yields,
\begin{eqnarray}\label{L4}
&&-\frac{\Omega^{(3)}_{\pm}}{M}\left(-\frac{2m_0eQF^{(3)}}{r}-2m_0E^{(3)}_NF^{(3)}\right) +\frac{F^{(3)}}{M r^2} \left(-\mu - \frac{2i\mu(1-\cos\theta)}{\sin^2\theta}\frac{\partial}{\partial\phi}-\frac{\mu^2(1-\cos\theta)^2}{\sin^2\theta}\right)\Omega^{(3)}_{\pm} \nonumber\\
&&+ \frac{1}{M^2}\vec{\nabla}M \cdot \vec{\nabla}\psi^{(3)}_{\pm}+  \left(\frac{1}{2M^2}\nabla^2M - \frac{3}{4M^3} \vec{\nabla}M\cdot\vec{\nabla}M\right) F^{(3)} \Omega^{(3)}_{\pm} = E_{N} F^{(3)} \Omega^{(3)}_{\pm}, 
\end{eqnarray}
the angular term containing the first derivative of $\Omega^{(3)}$ in $\phi$, which can be simply obtained from (\ref{harmonico}), namely,
\begin{equation}\label{azimutal}
\frac{\partial \Omega^{(3)}_{\pm} }{\partial \phi} = i(\mu + m \mp 1/2)\Omega^{(3)}_{\pm},
\end{equation}
cannot be dropped out because the coefficients $(\mu + m + 1/2)$ and $(\mu + m - 1/2)$ cannot be simultaneously null, that is, we could not consider an azimuthal symmetry in such system. Thus, replacing the result of this derivative in equation (\ref{L4}) and putting the product $F^{(3)}\Omega^{(3)}_{\pm}$ in evidence, we obtain,
\begin{eqnarray}\label{L5}
&&\left(\frac{2m_0eQ}{Mr}+\frac{2m_0E^{(3)}_N}{M} -\frac{\mu}{M r^2} + \frac{2\mu(1-\cos\theta)(\mu + m \mp 1/2)}{M r^2\sin^2\theta}-\frac{\mu^2(1-\cos\theta)^2}{M r^2\sin^2\theta}\right. \nonumber\\
&&\left.+\frac{1}{M^2\psi^{(3)}_{\pm}}\vec{\nabla}M \cdot \vec{\nabla}\psi^{(3)}_{\pm}+  \frac{1}{2M^2}\nabla^2M - \frac{3}{4M^3} \vec{\nabla}M\cdot\vec{\nabla}M\right) F^{(3)} \Omega^{(3)}_{\pm} = E_{N} F^{(3)} \Omega^{(3)}_{\pm}, 
\end{eqnarray}
eliminating $F^{(3)} \Omega ^{(3)}_{\pm}$, multiplying both the sides for $2M^2$, regrouping the terms and implementing the necessary simplifications, we get the following uncoupled system of Partial Differential Equations (PDE) \cite{SF},
\begin{eqnarray}\label{equationdyon3}
&&\nabla^2 M - \frac{3}{2}\frac{(\vec{\nabla}M)^2}{M}+\frac{4m_0 eQ M}{r} + \frac{2}{\psi_{\pm}^{(3)}}\vec{\nabla}M \cdot \vec{\nabla}\psi_{\pm}^{(3)}-\frac{2M\mu}{r^2}+\frac{4\mu(\mu + m \mp 1/2)(1-\cos\theta)M}{r^2{\sin}^2\theta}\nonumber\\
&&-\frac{2\mu^2(1-\cos\theta)^2M}{r^2{\sin}^2\theta}= 2M(M-2m_0)E_{N}^{(3)}.
\end{eqnarray}
Each PDE of the uncoupled system (\ref{equationdyon3}) can be considered as our mapping equation. Let us consider, for sake of simplicity, that the mass distribution $M$ depends only on the radial coordinate, namely $M=M(r)$, we need also to fix a particular value of $\theta$, say $\theta_0$. Thus, replacing $eQ$ by $n \alpha$ and considering that the charged spin-$1/2$ fermion to be mapped has constant mass $m_0=m_e=1$, the two equations (\ref{equationdyon3}) become Ordinary Differential Equations (ODE) \cite{SF}. These equations take the following form,
\begin{eqnarray}\label{equation radial dyon 3}
&&\frac{d^2M}{dr^2} +\frac{2}{r}\frac{dM}{dr}-  \frac{3}{2M}\left(\frac{dM}{dr}\right)^2 + \frac{2}{F_{k\ell}^{(3)}}\frac{dF_{k\ell}^{(3)}}{dr}\frac{dM}{dr}+\frac{4\mu(\mu + m \mp 1/2)(1-\cos\theta)M}{r^2{\sin}^2\theta}\nonumber\\
&& +\frac{4n \alpha M}{r}-\frac{2M\mu}{r^2}-\frac{2\mu^2(1-\cos\theta)^2M}{r^2{\sin}^2\theta}= 2M(M-2)E_{N}^{(3)},
\end{eqnarray}
where the term $4n\alpha M/r$ is related to the Coulomb electric potential .The system composed of a charged particle and a dyon has a discrete spectrum of energy, because of the electric charge $Q$ of dyon. Thus, the energy eigenvalues are obtained from,
\begin{equation}
E_N^{(3)}=-\frac{(n \alpha)^2}{2(N+1)^2},
\end{equation}
where $n$ is an integer and the quantum number $N = 0, 1, 2, 3, ...$. The equations (\ref{equation radial dyon 3}) are a generalization of the equations $(20)$ in reference \cite{monopole pdm pauli}.

\subsubsection{Numerical Solutions}

Considering that the mass depends only on the radial coordinate, namely $M=M(r)$, we need to choose a particular value for $\theta_0$, in order to solve the equations (\ref{equation radial dyon 3}) numerically. Let us consider a standard initial value problem (IVP), namely, $M(r_i) = 1$ and $M'(r_i) = 0$ for $r_i=0.20$ and let us fix $\theta_0 = 30$º. Due to the term containing the factor $(\mu + m \pm 1/2)$ --- one for each equation, only differing in the signal of $1/2$ ---  both equations of the system may have different solutions, which is not physically acceptable in the present situation, however we verify that both equations have approximately the same solution $M(r)$ in the interval $0.20 < r < 0.50$ if the following condition approximately holds,
\begin{equation}\label{solve dyon condition}
10 \leq|\mu + m|\leq 20,
\end{equation}
that is, for values of $\mu + m$ out of (\ref{solve dyon condition}) there is no mapping between the charge-dyon system --- considering low energy regime --- and some PDM quantum system. The equation (\ref{equation radial dyon 3}) will be numerically solved by the Mathematica software considering $\mu = 7$, $m = 13/2$, the ground state $N = 0$ and $n=-1800$ (bound state). The solutions for the two equations (\ref{equation radial dyon 3}) are presented in Fig. $1$. We can note that both solutions match in the interval $0.20 < r < 0.50$ and tend to infinity near $r_i$. The effective potential corresponding to this solution is given by expression (\ref{effective potential ZK}) and its curve is plotted in Fig. $2$.

In Fig. $3$ we present eigenvalues of energy $E_N^{(3)}$ for some values of $N$. We can observe that the curve moves to the origin when the eigenvalues of energy increase, overlapping for great values of $N$. The results obtained for the mass distribution in \cite{monopole pdm pauli} and those obtained here are compared in Fig. $4$. It is important to comment that it has been verified that there is no solutions of the equations (\ref{equation radial dyon 3}) for $n < - 3000$.

% Type 3 -dyon
\begin{figure}[!htp]
\centering
\includegraphics[scale=1.0]{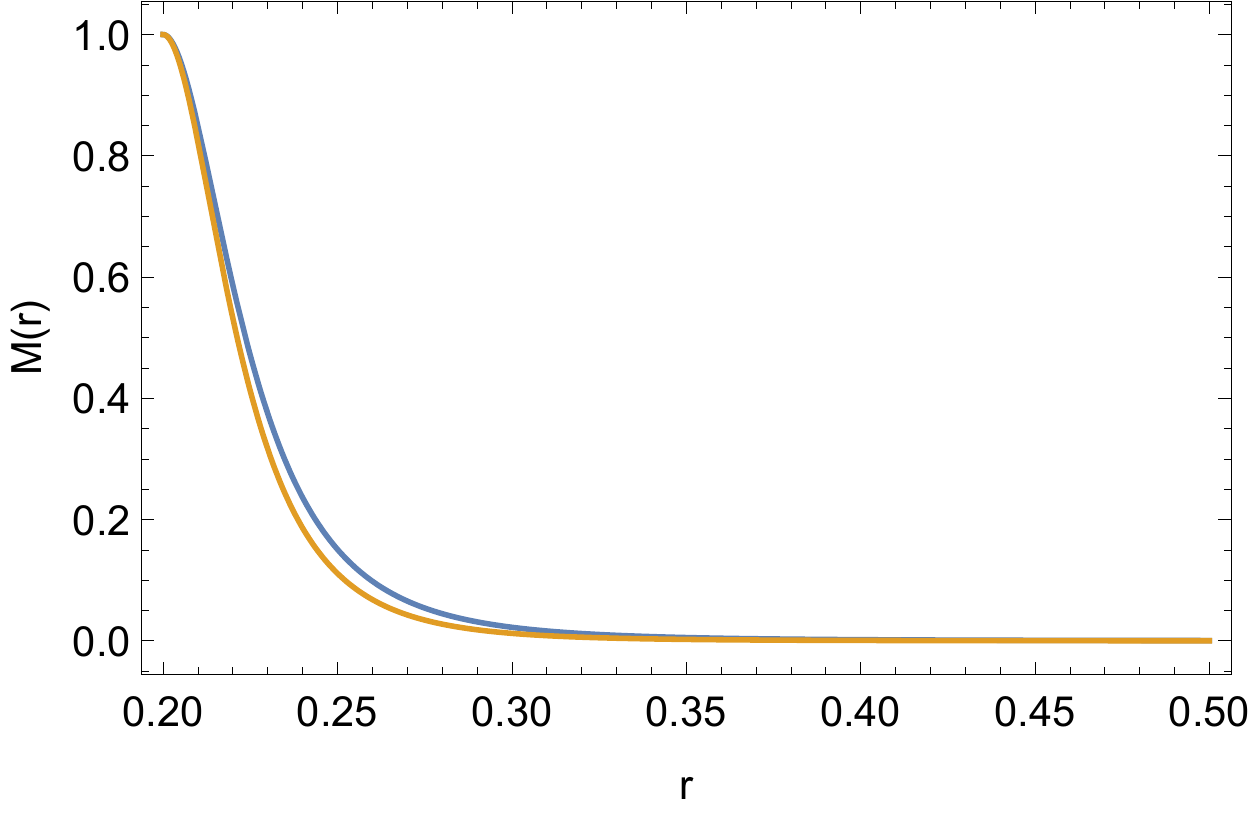}
\caption{Plot of numerical solutions of equation (\ref{equation radial dyon 3}), the dyon charge $Q$ was fixed in $-1800\sqrt{\alpha}$. Initial conditions are $M(r)$ using $M(r_i) = 1$ and $M'(r_i) = 0$ for $r_i=0.20$, here we fix $\theta_0 = 30$º. As close as one approaches the origin, where the dyon is located, the solution diverges. We use $\mu=7$, $j = m = 13/2$, $N = 0$ and units where $\hbar=m_e=c=1$.}
\end{figure}

%Type 3 - dyon
\begin{figure}[!htp]
\centering
\includegraphics[scale=1.0]{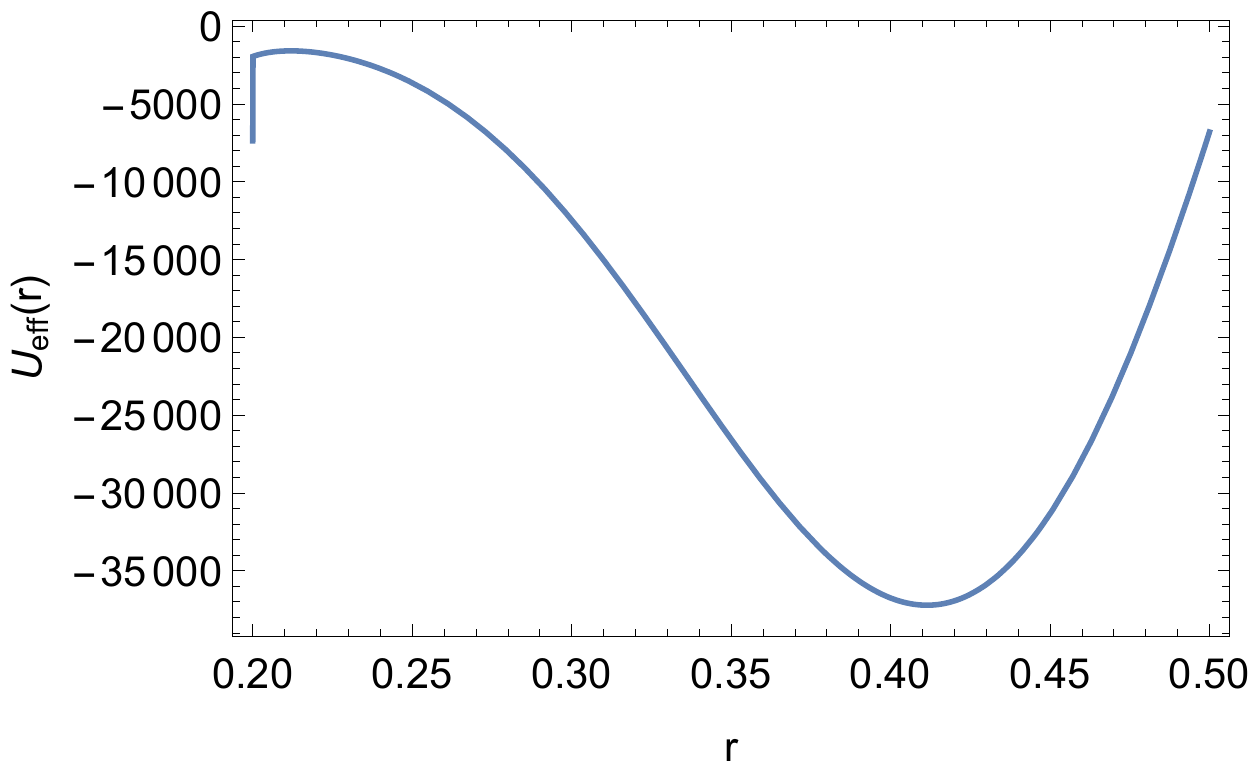}
\caption{Effective potential calculated using the solution of the system (\ref{equation radial dyon 3}). Boundary conditions are $M(r)$ using $M(r_i) = 1$ and $M'(r_i) = 0$ for $r_i=0.20$, here we fix $\theta_0 = 30$º. As close as one approaches the origin, where the dyon is located, the solution diverges. We use $\mu=7$, $j = m = 13/2$, $N = 0$ and units where $\hbar=m_e=c=1$.}
\end{figure}

Utilizing the Mustafa-Mazharimousavi ordering \cite{mustafa}, where $\alpha=\gamma = -1/4$, and $\beta= -1/2$, the effective potential (\ref{pot eff}) is written as,
\begin{equation}
U_{eff}=\frac{1}{4}\frac{\nabla^2 M}{M^2}-\frac{7}{16}\frac{(\vec{\nabla}M)^2}{M^3},
\end{equation}
however, replacing this expression in the equation (\ref{Model equation dyon}) and developing the calculations, the results are practically the same as those presented for the Zhu-Kroemer ordering .

%autovalores de energia
\begin{figure}[!htp]
\centering
\includegraphics[scale=1.2]{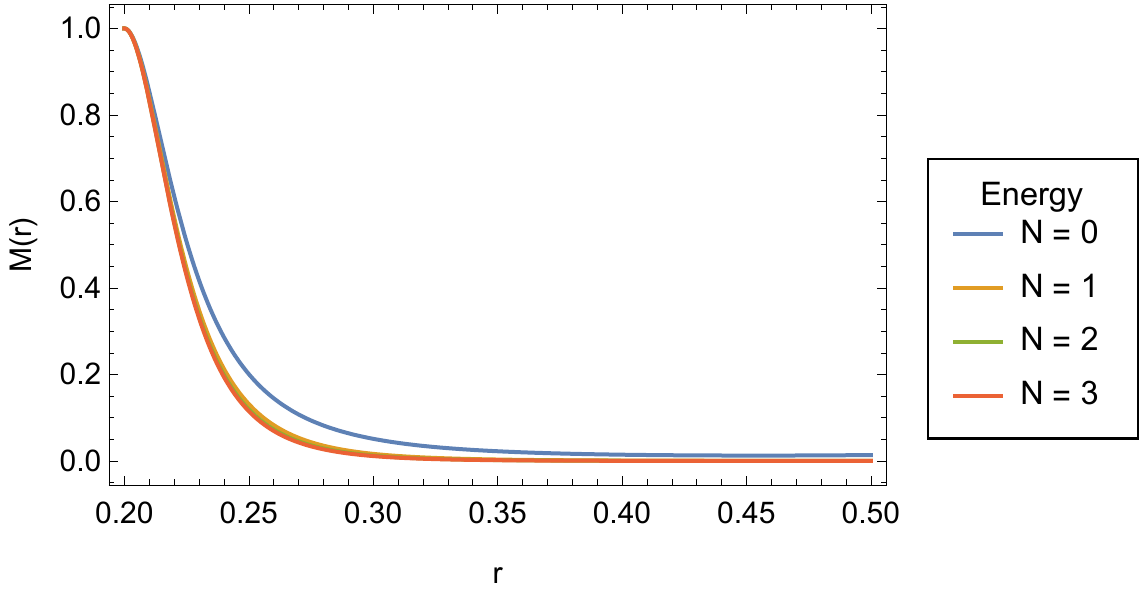}
\caption{Plot of numerical radial solutions related to some energy eigenvalues given by $N$. The curve moves to the origin when the eigenvalues increase. Boundary conditions are $M(r)$ using $M(r_i) = 1$ and $M'(r_i) = 0$ for $r_i=0.20$, here we fix $\theta_0 = 30$º and $n = - 2800$. We use $\mu=7$, $j = m = 13/2$ and units where $\hbar=m_e=c=1$.}
\end{figure}

%Comparação monopolo  e dyon - Type 3
\begin{figure}[!htp]
\centering
\includegraphics[scale=1.2]{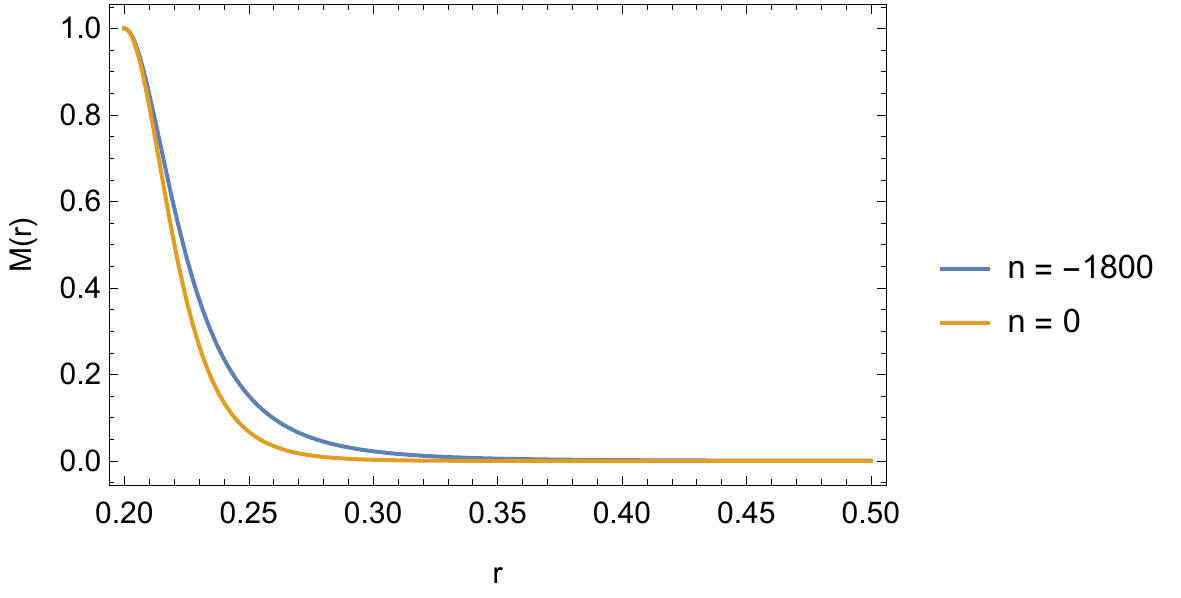}
\caption{Plot of numerical solutions for type 3. We compare the mass distribution considering two values of $n$. For $n=0$ we have a monopole at the origin.  Boundary conditions are $M(r)$ using $M(r_i) = 1$ and $M'(r_i) = 0$ for $r_i=0.20$, here we fix $\theta_0 = 30$º. As close as $n$ increase from $-1800$ to $0$, the curve approaches the origin. We use $\mu=7$, $j = m = 13/2$, $N=0$ and units where $\hbar=m_e=c=1$.}
\end{figure}

\section{Conclusion}

We studied the non-relativistic charge-dyon system in a position-dependent mass background via the Pauli equation and introduced a mapping between the former and a PDM quantum system. We derived the PDM free Pauli equation and replaced an exact solution (Pauli bi-spinor) for the charge-dyon system into this equation to obtain an uncoupled system with two non-linear partial differential equations (\ref{equationdyon3}) for the mass distribution.

We dealt with the case concerning the eigenvalues of the operator $\vec{J}$ starting from the minimal value $j = \mu-1/2$ (named type 3). We considered a radial dependence only, as well as the approximate condition (\ref{solve dyon condition}), which is necessary for the solution of both equations of the system (\ref{equation radial dyon 3}) to be approximately the same. We solved the equations numerically and plotted the results, as well as the graphics of the effective potentials associated (which represent our analogous models).

Thus we illustrate our results in figures 1-4. In Fig. $4$  we compare the mass distribution for the charge-monopole system $(n=0)$ and the charge-dyon system $(n=-1800)$. Finally, it is useful to remark that this approach can lead to a simulation of a charge-dyon system with spin $1/2$ in a non-relativistic case; in other words, to simulate a physical system not observed in nature yet.

Concluding, the technique developed throughout this work could serve as a basis to lead to an experimental model, maybe using a controlled system in condensed matter, which could be used to simulate some physical systems in the laboratory. Thus we can state that the effective potential, which was determined for each non-relativistic mapping, \textbf {provides a way to carry out the analogous models of the target systems presented in our work}. A possible suggestion for an experimental implementation could be the technique known as {\it Molecular Beam Epitaxy} (MBE) \cite {mbe, mbe2}, a powerful experimental technique used mainly for the growth of high quality semiconductor layers and film deposition thin and ultrafine \cite{fino, fino2}. This technique is used basically in the construction of devices and in basic research. The theoretical technique developed in this work could even lead to possible applications in the electronic device and data storage industry in the future.

\acknowledgments The authors gratefully acknowledge INCT-IQ and CNPq for partial financial support. This study was financed in part by the Coordena\c c\~ao de Aperfei\c coamento de Pessoal de N\'ivel Superior --- Brasil (CAPES) --- Finance Code 001.

\vspace{0.5cm}
\small{\textbf{Data Availability Statement}}
\vspace{0.2cm}

The data that support the findings of this study are available from the corresponding author upon reasonable request.

\end{document}